\pdfoutput=1
\documentclass[%
reprint,
superscriptaddress,
 amsmath,amssymb, 
 aps,
 prl,
floatfix,
]{revtex4-1}

\usepackage{graphicx}
\usepackage{epsfig} 
\usepackage{dcolumn}
\usepackage{bm}
\usepackage{epstopdf}
\usepackage{multirow}
\usepackage{amsmath}
\usepackage{booktabs}
\usepackage{color}
\usepackage{gensymb}
\usepackage{kantlipsum}  
\usepackage{hyperref}
\usepackage{ulem}
\usepackage{braket}
\usepackage{physics}
\usepackage[utf8]{inputenc}
\usepackage[T1]{fontenc}
\DeclareMathAlphabet{\altmathcal}{OMS}{cmsy}{m}{n}
\usepackage{mathptmx}
\usepackage[dvipsnames]{xcolor}
\usepackage{float}
\newcommand{\rom}[1]{\uppercase\expandafter{\romannumeral #1\relax}}
\newcommand*{\mr}{\mathrm} 

\begin{document}

\title{Observation of Self-Bound Droplets of Ultracold Dipolar Molecules}
\preprint{APS/123-QED}

\author{Siwei Zhang}
\thanks{These authors contributed equally.}
\affiliation{Department of Physics, Columbia University, New York, New York 10027, USA}
\author{Weijun Yuan}
\thanks{These authors contributed equally.}
\affiliation{Department of Physics, Columbia University, New York, New York 10027, USA}
\author{Niccol\`{o} Bigagli}
\affiliation{Department of Physics, Columbia University, New York, New York 10027, USA}
\author{Haneul Kwak}
\affiliation{Department of Physics, Columbia University, New York, New York 10027, USA}
\author{Tijs Karman}
\affiliation{Institute for Molecules and Materials, Radboud University, 6525 AJ Nijmegen, Netherlands}
\author{Ian Stevenson}
\affiliation{Department of Physics, Columbia University, New York, New York 10027, USA}
\author{Sebastian Will}\email{Corresponding author. Email: sebastian.will@columbia.edu}
\affiliation{Department of Physics, Columbia University, New York, New York 10027, USA}

\date{\today}

\begin{abstract} 
Ultracold gases of dipolar molecules have long been envisioned as a platform for the realization of novel quantum phases~\cite{goral2002quantum, micheli2006toolbox, buchler2007three, buchler2007strongly, cooper2009stable, Capogrosso2010, gorshkov2011tunable, syzranov2014spin, schmidt2022self}. Recent advances in collisional shielding~\cite{gorshkov2008suppression, karman2018microwave, matsuda2020resonant, anderegg2021observation}, protecting molecules from inelastic losses, have enabled the creation of degenerate Fermi gases~\cite{de2019degenerate, valtolina2020dipolar, schindewolf2022evaporation} and, more recently, Bose-Einstein condensation of dipolar molecules~\cite{bigagli2024observation}. However, the observation of quantum phases in ultracold molecular gases that are driven by dipole-dipole interactions has so far remained elusive. In this work, we report the formation of self-bound droplets and droplet arrays in an ultracold gas of strongly dipolar sodium–cesium molecules. Starting from a molecular Bose-Einstein condensate (BEC), microwave dressing fields are used to induce dipole-dipole interactions with controllable strength and anisotropy. By varying the speed at which interactions are induced, covering a dynamic range of four orders of magnitude, we prepare droplets under equilibrium and non-equilibrium conditions, observing a transition from robust one-dimensional (1D) arrays to fluctuating two-dimensional (2D) structures. The droplets exhibit densities up to 100 times higher than the initial BEC, reaching the strongly interacting regime, and suggesting the possibility of a quantum-liquid or crystalline state~\cite{gorshkov2008suppression, ciardi2025self}. This work establishes ultracold molecules as a system for the exploration of strongly dipolar quantum matter and opens the door to the realization of self-organized crystal phases~\cite{buchler2007strongly, gorshkov2008suppression, pupillo2008cold} and dipolar spin liquids in optical lattices~\cite{yao2018quantum}.
\end{abstract}

\maketitle

\section{Introduction}

The emergence of droplets is ubiquitous in nature, typically arising from a delicate interplay between attractive and repulsive forces, both in classical and quantum many-body systems. Examples include the formation of galaxies~\cite{chandrasekhar1953magnetic}, stars~\cite{kennicutt2012star}, water droplets~\cite{young1805iii, stillinger1980water}, atomic nuclei~\cite{gamow1930mass, weizsacker1935theorie, bohr1939mechanism}, and superfluid helium droplets~\cite{grebenev1998superfluidity, volovik2003universe}. In ultracold atomic quantum gases, droplet formation has been observed in bosonic mixtures with short-range van der Waals interactions~\cite{petrov2015quantum, cabrera2018quantum}, as well as in quantum gases of magnetic atoms with long-range dipolar interactions~\cite{schmitt2016self, chomaz2016quantum}. For magnetic atoms, complex droplet structures have been realized, including droplet arrays~\cite{kadau2016observing} and supersolid phases~\cite{tanzi2019observation, bottcher2019transient, chomaz2019long, norcia2021two}. In both platforms, the droplets are stabilized by repulsive forces originating from quantum fluctuations. Comparing the range of interactions to the interparticle spacing, droplets in atomic quantum gases have so far been dilute and weakly interacting.

Ultracold dipolar molecules are emerging as a powerful platform for many-body physics, quantum simulation, and quantum information~\cite{demille2002quantum, carr2009cold, baranov2012condensed, cornish2024quantum}. Based on the prospect of realizing many-body systems with strong dipolar interactions, theoretical efforts over the past two decades have pointed out their potential to realize novel quantum phases, such as quantum droplets~\cite{schmidt2022self}, supersolids~\cite{lechner2015tunable, lu2015stable, hertkorn2021pattern}, dipolar crystals~\cite{buchler2007strongly, gorshkov2008suppression} and strongly correlated lattice models~\cite{micheli2006toolbox, Capogrosso2010, gorshkov2011tunable}. On the experimental side, strong inelastic losses have long posed a major obstacle to creating quantum degenerate gases of molecules~\cite{ospelkaus2010quantum, bause2023ultracold}. Recent breakthroughs came with the development of collisional shielding techniques, using static and oscillating electric fields~\cite{matsuda2020resonant, anderegg2021observation, bigagli2023collisionally, lin2023microwave}, which have enabled the creation of degenerate Fermi gases~\cite{valtolina2020dipolar, schindewolf2022evaporation} and Bose-Einstein condensation~\cite{bigagli2024observation} of dipolar molecules in regimes where dipole-dipole interactions remain relatively weak. However, the creation of degenerate molecular gases with strong dipolar interactions has remained elusive, and it has been a critical open question whether they can be sufficiently long-lived to enable the observation of novel quantum phases.

\begin{figure*}[]
    \centering
    \includegraphics[width = 13.6 cm]{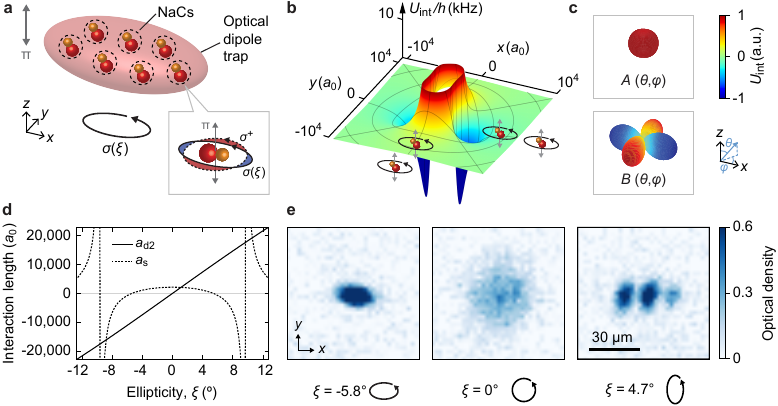}\\
    \caption{\textbf{Droplet formation in an ultracold gas of NaCs molecules with non-axially symmetric dipolar interactions.} \textbf{a}, A quantum degenerate gas of microwave-dressed NaCs molecules is confined in an anisotropic optical dipole trap. The molecules are simultaneously dressed with a $\pi$ field with linear polarization along the $z$ direction and a $\sigma$ field with elliptical polarization rotating in the $xy$ plane. The frequencies of the dressing fields are close to 3.5 GHz. By tuning the ellipticity $\xi$ of the $\sigma$ field away from pure $\sigma^+$ circular polarization, long-range interactions between the molecules take the form of a dipolar interaction for which the axial symmetry is broken. Larger ellipticity $\abs{\xi}$ leads to stronger dipolar interactions. Gravity points along the $-z$ direction. \textbf{b}, Intermolecular interaction potential in the $xy$ plane, calculated for $\xi = -6^\circ$. \textbf{c}, Angular dependence of the short-range part (top) and the long-range part (bottom) of the interaction in three-dimensional space for a fixed intermolecular distance ($\xi = -6^\circ$). \textbf{d}, Interaction length scales as a function of $\xi$, calculated from the microwave-dressed interaction potential. The solid (dashed) line shows the dipolar length $a_{\mr{d2}}$ (s-wave scattering length $a_{\mr{s}}$) as defined in the main text. \textbf{e}, Absorption images of a single droplet (left), a weakly dipolar BEC (middle), and a 1D-droplet array (right) after 25 ms time-of-flight expansion. Each image is an average of three individual shots taken along the $z$ direction. Black ellipses (bottom) indicate the trace of the electric field vector of the $\sigma$ field. Scale bar, 30 \textmu m (\textbf{e}).}
    \label{fig:1}
\end{figure*}

Here, we report on the observation of self-bound droplets in an ultracold gas of dipolar sodium–cesium (NaCs) molecules. Starting from a weakly interacting molecular BEC, we use microwave dressing fields to ramp up the interaction strength and observe the formation of both single droplets and droplet arrays. By varying the interaction ramp speed over four orders of magnitude, we observe a transition from stable 1D droplet arrays to fluctuating 2D structures, signaling a crossover from near-adiabatic to non-adiabatic state preparation. In the single-droplet phase, we measure an increase of the molecular density by two orders of magnitude compared to the initial BEC. In this regime, the characteristic range of dipolar interactions significantly exceeds the interparticle spacing, indicating the realization of a dipolar many-body system in the strongly interacting regime. 

\section{Strongly Dipolar Interactions}

We control the interactions and collisional properties in our molecular samples using two microwave fields that dress the molecules' first rotational transitions~\cite{karman2025double, yuan2025upcoming}, as illustrated in Fig.~\ref{fig:1}. The two microwave fields are elliptically-polarized rotating in the $xy$ plane ($\sigma$ field) and linearly-polarized oscillating along the $z$ axis ($\pi$ field), respectively, see Fig.~\ref{fig:1}\textbf{a}. As the molecular dipoles follow the oscillating electric fields, a time-averaged  interaction potential arises that can be well approximated by the sum of a repulsive van der Waals ($\propto 1/r^6$) and a dipolar long-range ($\propto 1/r^3$) term \cite{karman2025double}, see Fig.~\ref{fig:1}\textbf{b}:
\begin{equation}
\nonumber
U_{\mr{int}}(\vec{r}) = \frac{A(\theta, \varphi)}{r^\mr{6}} +  \frac{B(\theta, \varphi)}{r^\mr{3}}.
\end{equation}
Here, $r$, $\theta$, and $\phi$ define the relative position vector between two molecules in spherical coordinates, as shown in Fig.~\ref{fig:1}\textbf{c}. 

The van der Waals contribution constitutes a strongly repulsive hardcore potential for intermolecular distances $r \lesssim 2,000 \ a_0$. The $A(\theta, \varphi)$ term is positive for all angles $\theta$ and $\varphi$ and nearly isotropic (see Methods). The hardcore repulsion shields the molecules against fast inelastic losses and ensures the collisional stability of the molecular gas~\cite{gorshkov2008suppression, karman2018microwave}. In addition, the hardcore potential sets an upper limit for the density that can be reached in the molecular sample. 

\begin{figure*}[t]
    \centering
    \includegraphics[width = 18 cm]{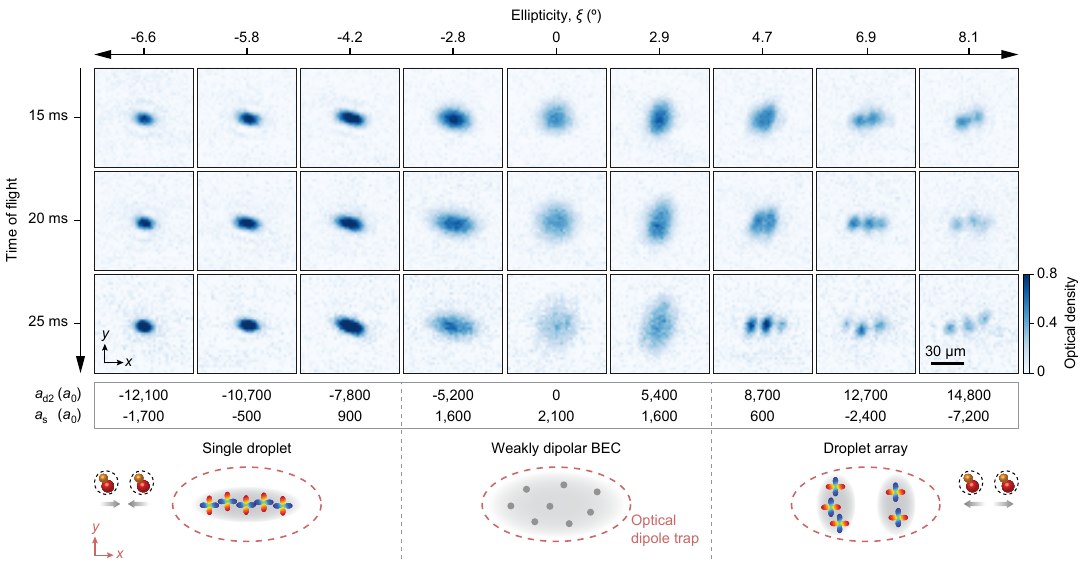}\\
    \caption{\textbf{Single droplets and droplet arrays.} After ramping the $\sigma$ field ellipticity to the target value $\xi$ within 200 ms, absorption images after 15, 20, and 25 ms time of flight are shown in the top, middle, and bottom row, respectively. Each image is an average of 3 individual shots, taken along the $z$ direction. The molecular clouds do not show a discernible thermal component due to the high condensate fraction. Below each column, the respective values of $a_\mathrm{d2}$ and $a_\mathrm{s}$ are shown. (bottom) Illustration of the orientation of the non-axially symmetric dipolar interaction potential with respect to the trap axes and the inferred in-trap density profile of the molecular cloud (gray shading). Scale bar, 30 \textmu m.} 
    \label{fig:2}
\end{figure*}

The strength and angular dependence of the dipolar contribution is controlled by the microwave parameters as follows: 
For a perfectly circularly polarized $\sigma$ field (ellipticity $\xi = 0\degree$), the strength of the $\pi$ field can be tuned such that dipolar interactions vanish at long range ($B = 0$) as the dipolar interactions induced by the $\sigma$ and $\pi$ fields compensate each other~\cite{yuan2025upcoming}. Under these conditions, where the intermolecular potential is dominated by the hardcore van der Waals potential, we can efficiently cool molecular samples into a BEC, as described in Ref.~\cite{bigagli2024observation}. While keeping the $\pi$ field fixed at this compensation point, changing the ellipticity of the $\sigma$ field induces a dipolar interaction in which the axial symmetry is broken \cite{syzranov2014spin}; its anisotropy is illustrated in  Fig.~\ref{fig:1}\textbf{c} and described by~\cite{chen2023field}
\begin{equation}
\nonumber
B(\theta, \varphi) = a_{\mr{d2}}(\xi) \frac{\hbar^2}{m} {\mr{\sqrt{3} \cos 2 \varphi \sin ^2 \theta}},
\end{equation}
where $m$ is the mass of the molecule and $\hbar$ the reduced Planck constant. The strength of the dipolar interactions is parametrized by the dipolar length, $a_{\mr{d2}}(\xi) \propto \sin2\xi$. Larger ellipticity leads to stronger dipolar interactions. The sign of $\xi$ encodes the orientation of the polarization ellipse, with positive (negative) values indicating alignment along the $y$ ($x$) axis. The angular anisotropy of $B(\theta, \varphi)$ corresponds to the spherical harmonic $d_{x^2 - y^2}$, which has no axial symmetry along any axis. This form of dipolar interactions is distinctly different from the conventional $d_{3 z^2 - r^2}$ form, which is axially symmetric about the $z$ axis and widely used in ultracold gases of magnetic atoms~\cite{chomaz2022dipolar}. In addition to tuning the ellipticity, there is further flexibility in controlling the anisotropy by tuning the $\pi$ field away from the compensation point, allowing the continuous tuning between axially symmetric and non-axially symmetric dipolar interactions. 

Our experiments start with the preparation of a BEC of NaCs molecules in their rovibrational ground state. The molecular gas is confined in an anisotropic optical dipole trap with frequencies \{$\omega_x, \omega_y, \omega_z$\} = 2$\pi\times$\{15(2), 27(2), 53(4)\}~Hz, after evaporative cooling from about 700 nK to less than 10 nK in the presence of $\sigma$ and $\pi$ dressing fields  (see Methods). At this stage, the ellipticity of the $\sigma$ field is $\xi = 0(1)\degree$, resulting in a gas of hardcore bosons with minimal dipolar interactions. The BEC contains 1,500(300) molecules, with a condensate fraction greater than $70 \%$. Then, we induce dipolar interactions by linearly ramping the ellipticity $\xi$ to its target value within 200 ms. The values of the dipolar length $a_\mathrm{d2}$ and the s-wave scattering length $a_\mathrm{s}$ as a function of $\xi$ are shown in Fig.~\ref{fig:1}\textbf{d}. The s-wave scattering length $a_\mathrm{s}$ decreases from 2,100 $a_0$ at $\xi = 0\degree$ to negative infinity at $\xi = \pm 9.7\degree$, where the appearance of a bound state in the intermolecular potential gives rise to scattering resonances~\cite{lassabliere2018controlling, chen2023field}. Between these resonances, the dipolar length $a_\mathrm{d2}$ can be continuously tuned from 0 to $\pm 18,000$ $a_0$, where the negative (positive) sign indicates that the molecules attract (repel) each other along the $x$ axis and repel (attract) each other along the $y$ axis. By tuning the interactions across this range, we observe strikingly different phases of the degenerate molecular gas. Fig.~\ref{fig:1}\textbf{e} shows absorption images for $\xi = 0(1)^\circ$, $-5.8(5)^\circ$, and $4.7(5)^\circ$ after 25 ms time-of-flight expansion, corresponding to a weakly dipolar BEC, a single droplet, and a droplet array, respectively.

\begin{figure*} [t]
    \centering
    \includegraphics[width = 18.0 cm]{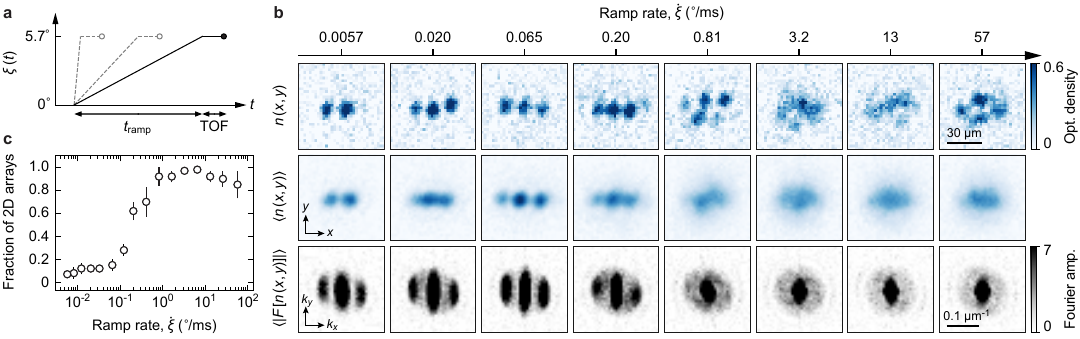}\\
    \caption{\textbf{Droplet arrays for slow and fast interaction ramps.} \textbf{a}, Experimental sequence. The ellipticity is linearly increased to $5.7^\circ$ while the ramp time $t_\mathrm{ramp}$ and the associated ramp rate $\dot{\xi}$ are varied. Absorption images of the resulting droplet arrays are recorded after 25 ms time of flight (TOF). \textbf{b}, For various ramp rates $\dot{\xi}$, typical single-shot images (top), averages of 30 individual shots (middle), and averages of the modulus of their Fourier transforms (bottom) are shown. \textbf{c}, Fraction of images showing 2D arrays for a given ramp rate $\dot{\xi}$. Each data point is obtained by counting the fraction of shots (out of a sample of 30 shots) that exhibit 2D arrays. Error bars account for shots for which the array dimensionality is ambiguous. Scale bars, 30 \textmu m for absorption images and 0.1 \textmu m$^{-1}$ for Fourier transforms (\textbf{b}).
    } 
    \label{fig:3} 
\end{figure*}

\section{Expansion Dynamics}

We distinguish the different phases of the molecular gas by analyzing its expansion dynamics after release from the optical dipole trap. Fig.~\ref{fig:2} shows absorption images for various ellipticities $\xi$ after 15, 20, and 25 ms time of flight. While for each of the shown ellipticities there are noticeable differences in the expansion behavior, we phenomenologically group the observed states into three distinct phases:

For ellipticities between $-3.5(5)^\circ$ and $3.5(5)^\circ$, the expansion dynamics are consistent with a weakly dipolar BEC~\cite{stuhler2005observation}. At $\xi = 0(1) \degree$, where dipolar interactions are minimized, the gas expands into an approximately round cloud, as expected for a non-dipolar, repulsively interacting BEC for our trap parameters at the specific expansion times \cite{ketterle1999making}. As $\xi$ deviates from zero, we observe electrostriction: the clouds show a significant elongation along the axis of attractive dipolar interactions. 

At stronger dipolar interactions, $|\xi| > 3.5^\circ$, we observe the formation of molecular droplets. The behavior differs markedly for the two orientations of the microwave polarization ellipse, as a result of the interplay between the alignment of the dipolar interaction pattern and the trap anisotropy. For positive (negative) ellipticity the microwave polarization ellipse is aligned along the tighter (looser) trap direction in the $xy$ plane.

For negative ellipticities, $\xi < -3.5^\circ$, the gas forms a single self-bound droplet. Compared to the weakly dipolar BEC, the cloud is significantly contracted and shows no observable expansion as a function of time of flight. Taking the finite imaging resolution (see Methods) into account, we infer that the expansion energy is at least 20 times smaller than for the weakly dipolar BEC. Additional images taken along the $x$ axis, see Fig.~\ref{fig:SI6}, confirm that the droplet is self-bound in all directions. For positive ellipticities, $\xi > 3.5^\circ$, the gas initially develops a density modulation, which evolves into an array of self-bound droplets at higher ellipticities. During time of flight, the individual droplets do not expand, but move apart from each other due to repulsive side-by-side dipolar interactions~\cite{ferrier2016observation}. A quantitative analysis of the cloud sizes during time-of-flight expansion, shown in Fig.~\ref{fig:SI1}, corroborates the self-bound character of both single-droplet and droplet-array phases.

\section{Adiabaticity of Interaction Ramp}

\begin{figure}[]
    \centering
    \includegraphics[width = 8.6 cm]{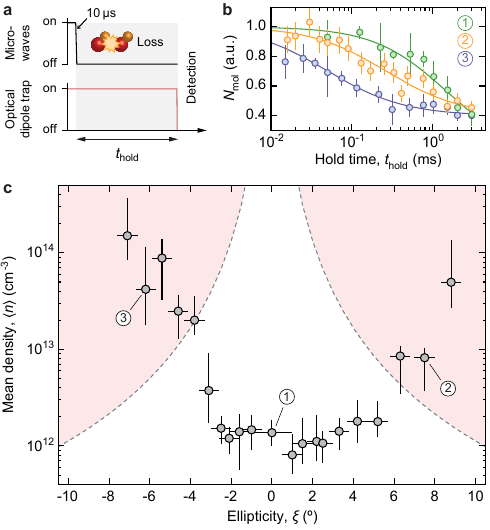}\\
    \caption{\textbf{Molecular density in the BEC and droplet phases.} \textbf{a}, Illustration of the experimental sequence. Microwave dressing is turned off within 10 \textmu s, inducing fast near-universal two-body loss. After a variable hold time $t_{\mr{hold}}$, the molecule number is measured via \textit{in situ} absorption imaging. \textbf{b}, Two-body decay curves for samples prepared at $\sigma$ field ellipticities  $-6.2^\circ$ (blue), $0^\circ$ (green), and $7.5^\circ$ (orange), corresponding to a single droplet, a weakly dipolar BEC, and a droplet array, respectively. Data points correspond to averages of 9 experimental runs; error bars show the uncertainty of the fit to the averaged cloud profile. Solid lines show the fit of a two-body loss model to extract the mean density at $t_{\mr{hold}}=0$ (see Methods). \textbf{c}, Mean molecular density as a function of ellipticity. Encircled numbers indicate the respective data in panel \textbf{b}. The pink shaded areas highlight mean densities for which $\left< n \right> \abs{a_{\mr{d2}}}^3 > 1$, indicating the strongly interacting regime where dipolar interaction energy dominates over kinetic energy in a degenerate gas. Horizontal error bars show the uncertainty in the ellipticity calibration; vertical error bars show the uncertainty in the fitted density.
    } 
    \label{fig:4}
\end{figure}

Next, we investigate the formation of droplet arrays as a function of interaction ramp speed, assessing to which extent the emergence of arrays is a ground-state phenomenon~\cite{bisset2016ground, ferrier2018onset}. As shown in Fig.~\ref{fig:3}\textbf{a}, the ellipticity is ramped linearly from $0^\circ$ to $5.7^\circ$. With ramp times between $100$ \textmu s and 1 s, the ramp rates $\dot{\xi}$ cover a dynamic range of four orders of magnitude. Fig.~\ref{fig:3}\textbf{b} shows typical single-shot absorption images after 25 ms time of flight, the average of several individual shots, and the averaged modulus of their Fourier transforms. The latter reveals the spectrum of spatial frequencies and the repeatability of the observed array structures. A comprehensive set of raw images is shown in Fig.~\ref{fig:SI2}.

For slow ramps, $\dot{\xi} < 0.1\degree$/ms, we consistently observe 1D arrays with two to four droplets aligned along the $x$ axis. The spacing between the droplets is stable from shot to shot, as evidenced by the modulation in the averaged Fourier transforms.
To confirm the near-equilibrium character of this state, we perform direct evaporative cooling to quantum degeneracy directly at $\xi = +5.7^\circ$. We observe arrays that closely match those obtained for slow ramps, suggesting that the 1D droplet array corresponds to the ground state or a low-lying metastable state of the interacting many-body system.

For faster ramps, $\dot{\xi} > 0.8\degree$/ms, we observe the emergence of 2D droplet arrays. The number and positions of droplets show strong shot-to-shot fluctuations. In the averaged images the density modulations wash out, reflecting the random character of the fluctuations. This is further evidenced by the absence of higher spatial frequency peaks in the averaged Fourier transforms. Notably, the observed density modulations of the 2D arrays occur both along the directions of repulsive and attractive dipolar interactions, suggesting a mechanism distinct from the roton instability discussed in quantum gases of magnetic atoms~\cite{chomaz2022dipolar}.

To identify the crossover between near-adiabatic and non-adiabatic preparation, we determine the fraction of shots showing 2D arrays for each $\dot{\xi}$. As shown in Fig.~\ref{fig:3}\textbf{c}, the transition occurs between $0.12\degree$/ms and $0.41\degree$/ms, corresponding to a ramp time of $t_\mathrm{ramp} \sim 20$ ms. This is also confirmed by an analysis of the aspect ratio of the clouds in the averaged images, see Fig.~\ref{fig:SI7}. These results confirm that the ramp rates used in Fig.~\ref{fig:2} ($\dot{\xi} < 0.04\degree$/ms) are at least near-adiabatic. 

\section{Droplet Density}

Given the strong contraction of the molecular clouds in the droplet phase, it is of particular interest to measure their density (see Fig.~\ref{fig:4}). Direct \textit{in situ} imaging is not feasible, as the droplet size is below our imaging resolution. Instead, we leverage the fast switchability of the microwave fields and utilize the density-dependence of two-body loss ($\propto n$) to infer the molecular density. After preparing the sample at a given ellipticity $\xi$, we turn off the dressing fields within 10 \textmu s (see Fig.~\ref{fig:4}\textbf{a}), initiating rapid two-body loss of the unshielded molecules \cite{ospelkaus2010quantum}. The molecular decay dynamics is observed for up to 3 ms; Fig.~\ref{fig:4}\textbf{b} shows examples for the BEC, droplet, and droplet array phases. In both droplet regimes, the decay is significantly faster than in the weakly dipolar BEC. Assuming that many-body correlations do not significantly modify the loss dynamics, enhanced decay corresponds to a higher molecule density~\cite{chomaz2016quantum, tanzi2019observation}. Fitting the observed decay curves using a numerical model based on a local density approximation (see Methods), we obtain the mean density $\langle n \rangle$ for each many-body state, as shown in Fig.~\ref{fig:4}\textbf{c}.

In the weakly dipolar BEC phase, $\abs{\xi}$ < $3.5\degree$, we measure a mean density of $10^{12}\, \mr{cm}^{-3}$, consistent with previous results~\cite{bigagli2024observation}. For $\xi < -3.5\degree$, the density increases sharply, possibly indicating a phase transition between the weakly dipolar BEC and the single droplet phase. For $\xi > 3.5\degree$, the density also rises as the system enters the droplet array phase, but the change is more gradual. In the single droplet phase, we measure mean densities of up to $\sim 10^{14}\,\mr{cm}^{-3}$, an increase by two orders of magnitude compared to the BEC, reaching values that are comparable to densities in ultracold atomic gases. At $\xi = -7.1(5)^\circ$, the measured mean interparticle spacing is $3,600(900)\,a_0$, very close to the lower bound set by the radius of the repulsive hardcore potential, $\sim 2,000\,a_0$. This raises the intriguing question whether the droplet approaches a quantum-liquid or crystalline state~\cite{gorshkov2008suppression, ciardi2025self}. The comparison of the dipolar length, $13,000(900)\, a_0$, with the interparticle spacing yields a ratio of 4(1), indicative of a system in the strongly interacting regime.  

Finally, we measure the lifetimes for all observed phases (see Fig.~\ref{fig:SI8}). Even for the densest and most strongly interacting states, the lifetimes exceed 100 ms, which is more than two orders of magnitude longer than the timescale of dipolar interactions, $\sim 1$ ms. This is a remarkable difference to strongly interacting atomic BECs where a large s-wave scattering length in the vicinity of a Feshbach resonance leads to fast recombination losses, limiting the lifetime to less than 1 ms~\cite{makotyn2014universal}. In microwave-dressed molecular samples recombination losses are absent \cite{yuan2025upcoming}, highlighting their potential for realizing strongly interacting quantum phases under equilibrium conditions. 

\section{Conclusions}

In conclusion, we have observed self-bound droplets and droplet arrays in a strongly dipolar quantum gas of ultracold molecules. Beyond the expansion dynamics and mean densities discussed here, further investigation of the observed phases will be of great interest. For the strongly interacting single droplet, measurements of vorticity, collective excitations, or shear response could clarify whether the phase has quantum-liquid or crystalline properties~\cite{gorshkov2008suppression, cinti2017classical, schmidt2022self, zhang2025quantum, ciardi2025self}. For the droplet arrays, probing phase coherence will elucidate whether the system can show supersolid behavior~\cite{tanzi2019observation, bottcher2019transient, chomaz2019long, norcia2021two, zhang2025supersolid}. Additional experimental and theoretical studies will be needed to understand the nature of the transitions between the observed phases and the stabilization mechanism of strongly dipolar droplets of microwave-dressed molecules~\cite{langen2025dipolar}.

This work significantly advances ultracold dipolar molecules as a platform for many-body quantum physics, quantum simulation, and the exploration of self-organization processes in the quantum regime. Microwave dressing provides highly tunable, long-range dipole-dipole interactions, in addition to a hardcore potential which ensures that even dense molecular samples are long-lived. The fast control of microwave fields further opens up new measurement modalities including quenches, non-equilibrium dynamics, and Floquet engineering of interactions. Beyond the 3D bulk systems explored here, exotic forms of self-organization are also expected in lower dimensions~\cite{wang2006quantum, buchler2007strongly, tang2018thermalization} and lattice geometries~\cite{micheli2006toolbox, Capogrosso2010, yao2018quantum}. Our system bridges a gap between well-understood model systems, such as ultracold atoms with contact s-wave interactions, and strongly interacting dense systems, such as superfluid helium~\cite{dalfovo2001helium}, electron gases~\cite{tsui1982two}, and nuclear matter~\cite{hammer2013colloquium}.

\section{Acknowledgements}

We acknowledge helpful discussions with Gordon Baym, Blair Blakie, Francesca Ferlaino, Tin-Lun Ho, Tim Langen, Andrew Millis, Tilman Pfau, Dam Son, Frank Wilczek, Wilhelm Zwerger, and Martin Zwierlein. We thank Lin Su and Tarik Yefsah for critical reading of the manuscript. We are grateful to Aden Lam and Claire Warner for important contributions in the construction of the experimental apparatus. This work was supported by an NSF CAREER Award (Award No.~1848466), an NSF Single Investigator Award (Award No.~2409747), an ONR DURIP Award (Award No.~N00014-21-1-2721), an AFOSR Single Investigator Award (Award No.~FA9550-25-1-0048), and a grant from the Gordon and Betty Moore Foundation (Award No.~GBMF12340). I.S.~was supported by the Ernest Kempton Adams Fund. T.K.~acknowledges NWO VIDI (Grant ID 10.61686/AKJWK33335). S.W.~acknowledges additional support from the Alfred P. Sloan Foundation.

%


\thispagestyle{empty}
\clearpage

\setcounter{figure}{0}
\makeatletter 
\renewcommand{\thefigure}{Extended Data \@arabic\c@figure}
\makeatother

\newpage

\section{Methods}

{\bf Sample preparation and detection.} Initially, a weakly interacting BEC is prepared via forced evaporation of 50,000 ground-state NaCs molecules~\cite{stevenson2023ultracold, cairncross2021assembly} from a temperature of 700(50) nK. The molecules are collisionally shielded via microwave dressing on the $J = 0$ to $J = 1$ rotational transition at 3.471 GHz, as described in previous work~\cite{bigagli2024observation}. The BEC is held in a crossed optical dipole trap with trap frequencies $\omega / (2 \pi) = \{15, 27, 53\}$~Hz. Both trapping laser beams are generated from a 1064 nm narrow-line single-mode Nd:YAG laser (Coherent Mephisto MOPA). The $x$-dipole trap is focused to waists of 172(1)~\textmu m (horizontal) and 54(1)~\textmu m (vertical), with the polarization set to linear along the $z$ direction; the $y$-dipole trap has waists of 210(1)~\textmu m (horizontal) and 58(1)~\textmu m (vertical), and is set to linear along the $x$ direction. Then, dipolar interactions are induced by tuning the microwave dressing fields as described in the main text.

The molecular samples are detected after turning off the trapping light and letting the molecules freely expand during time of flight. At the end of time of flight, the dressing fields are ramped down in 10~\textmu s, followed by fast (40~\textmu s) bound-to-free stimulated Raman adiabatic passage (STIRAP) to dissociate the molecules to Na and Cs atoms. Then, the Cs atoms are detected via standard optical pumping (75~\textmu s) and on-resonant absorption imaging (50~\textmu s). Additional details are provided in Ref.~\cite{bigagli2024observation}. The imaging resolution in our system is 3.8(3) \textmu m (1/$\sqrt{e}$ Gaussian radius). The STIRAP efficiency is 72(2)$\%$ at short time of flight and gradually decreases to 43(5)$\%$ at 25~ms time of flight. This change is taken into account in the molecular number calibration and the optical density of the image. When using \textit{in situ} imaging to measure the density, we decrease the intensity of the up-leg STIRAP light to lower the reverse STIRAP efficiency to 25(3)$\%$, which suppresses saturation effects and allows the measurement of density with higher accuracy.

\begin{figure*}[]
  \centering
  \includegraphics[width = 16.0 cm]{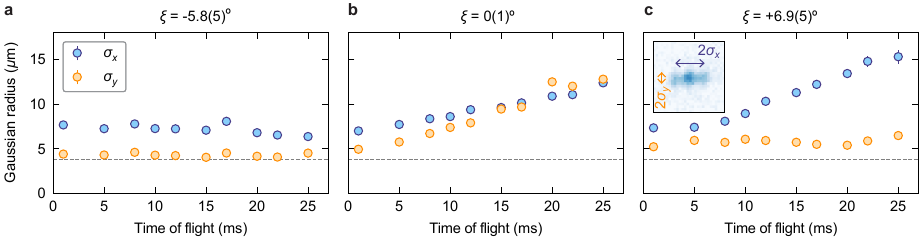}\\
  \caption{\textbf{Cloud size after time-of-flight expansion for the three different phases.} Measured Gaussian $1/e^{1/2}$ radius of the molecular gas as a function of time of flight for \textbf{a} single droplet, \textbf{b} weakly dipolar BEC, and \textbf{c} droplet array. Each data point is obtained from a Gaussian fit to an average of 15 shots, as shown in the inset of panel \textbf{c}. Error bars show the uncertainty of the fit to the averaged cloud profile. The gray dashed line indicates the imaging resolution of our system at 3.8(3) \textmu m.}
  \label{fig:SI1}
\end{figure*}

{\bf Image averaging procedure.}
When averaging images from multiple experimental runs, each image is re-centered to avoid unintended blur in the averaged image. When the molecular cloud is released from the trap for time-of-flight expansion, we observe a small center-of-mass velocity in random directions, likely due to a mechanical instability in our dipole trap setup. For BEC and single droplet images where no density modulation is present, we first perform a Gaussian fit on each image to determine the center-of-mass position. We then align and average the images according to their centers. For the droplet array images in Fig.~\ref{fig:3} and Fig.~\ref{fig:SI1}, the averaged images are obtained with the same centering procedure. For the droplet array images in Fig.~\ref{fig:1} and Fig.~\ref{fig:2} we apply a LoG (Laplacian of Gaussian) filter, extract the locations of the local maxima (droplets) in each image, and then align the images to the center position of the leftmost droplet.

\begin{figure}
    \centering
    \includegraphics[width = 8.6 cm]{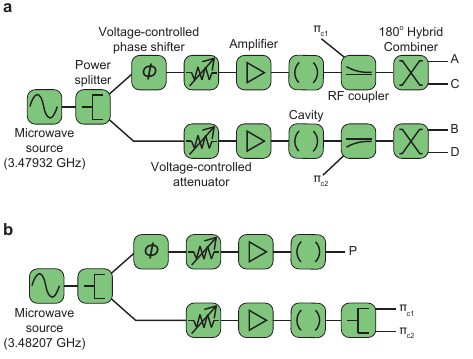}
    \caption{\textbf{Microwave setup.} \textbf{a}, Block diagram of the components generating the $\sigma$ field and the $\pi$ compensation field. The outputs A, B, C, and D each connect to the feed of a single loop antenna, together forming the cloverleaf antenna, positioned above the vacuum chamber. \textbf{b}, Block diagram of the components  generating the $\pi$ field. The outputs $\pi_\mathrm{c1}$ and $\pi_\mathrm{c2}$ are connected to the $\sigma$ field cloverleaf antenna via directional couplers. The output P connects to a single loop antenna, positioned on the side of the vacuum chamber. The microwave fields from the P antenna and the $\pi_\mathrm{c1}$ and $\pi_\mathrm{c2}$ components from the A, B, C, and D antennas coherently add together, in combination generating the $\pi$ dressing field.}
    \label{fig:SI5}
\end{figure}

{\bf Microwave setup.}
A simplified block diagram of the microwave system generating the microwave dressing fields is shown in Fig.~\ref{fig:SI5}. Omitted are microwave isolators, which are essential to protect the amplifiers from cavity-back reflections and to decouple the individual microwave chains.

The $\sigma$ microwave system, see Fig.~\ref{fig:SI5}\textbf{a}, feeds a cloverleaf-antenna array consisting of four loop antennas~\cite{yuan2023planar}. In the array, opposite antenna pairs, labeled A-C and B-D, generate orthogonal and almost-linear polarization, $E_{x'} \hat{x'}$ and $E_{y'} \hat{y'}$ respectively. Here, $\hat{x'}$ and $\hat{y'}$ denote unit vectors of a coordinate system that is slightly tilted with respect to the lab frame coordinates, $\hat{x}$ and $\hat{y}$, defined by the trap axes. The tilts are determined experimentally, see Section {\bf Microwave calibration}. To control $E_{x'}$, $E_{y'}$ and their relative phase $\phi$ independently, we use voltage-controlled attenuators (General Microwave D1954) both in the A-C and B-D paths, one voltage-controlled phase shifter (Sigatek SF40A2) in the A-C path, along with a 20~dB directional coupler after the cavity (not shown, Mini-Circuits ZUDC20-0283-S+) to pick off a portion of the signal for active stabilization and control. In the stabilization chain, we have a power detector (RF Bay RPD-5501) and a homemade phase detector. The phase detector consists of an I/Q mixer (Analog Devices HMC8193) fed by the pickup signals from A-C and B-D to detect the relative phase and an analog divider to turn the output of the I and Q ports of the mixer into a signal $\propto \cot{\phi}$. Error signals are formed with respect to analog set points. Feedback is applied via the phase shifter and attenuators through PI filters with $\sim 2~$kHz and $\sim 100~$kHz bandwidth, respectively. As the attenuators introduce phase shifts when attenuating the signal, phase stabilization is essential. The large separation in loop bandwidth is required to prevent oscillation. 

During the interaction ramp, the ellipticity of the $\sigma$ microwave field is controlled by tuning the relative power of the A-C and B-D microwave components ($E_{x'} \hat{x'}$ and $E_{y'} \hat{y'}$), while keeping the total Rabi frequency constant and the relative phase fixed at about $90\degree$. As a result, the orientation of the ellipse is always aligned along either $\hat{x'}$ or $\hat{y'}$, which are rotated by approximately $90\degree$ with respect to each other. 

The $\pi$ microwave system, see Fig.~\ref{fig:SI5}\textbf{b}, drives a single loop antenna (P) radiating mostly vertically polarized microwaves, and also generates compensation fields that are applied through the cloverleaf antenna array (A-C and B-D). 
The combination of two antennas is necessary to achieve a linear $\pi$ microwave that is perpendicular to the plane of the $\sigma$ microwave. The main $\pi$ microwave branch, connected to antenna P, produces microwaves with a Rabi frequency of $2\pi \times 7.2(1)$~MHz. The two compensation branches, denoted $\pi_{c1}$ and $\pi_{c2}$, are coupled into the cloverleaf antenna's A-C and B-D arms with directional couplers (Mini-Circuits ZUDC10-0283-S+). The compensation fields are added coherently to tilt the main $\pi$ field so that the projection of the total $\pi$ field onto the $\hat{x'} \hat{y'}$ plane is zero. The relative phase and power of $\pi_{c1}$ and $\pi_{c2}$ are controlled by discrete attenuators and cable delays. The main $\pi$ and compensation branches have independent power control loops along with a control loop that stabilizes their relative phase, identical to the $\sigma$ system. The compensation field has a coupling strength of about $2\pi \times 1.1(1)$~MHz while the total $\pi$ microwave Rabi frequency is $2\pi \times 7.1(1)$~MHz.

{\bf Microwave calibration.}
Following the approach discussed in Ref.~\cite{zhang2024dressed}, we use dressed-state spectroscopy on molecules to calibrate the Rabi frequencies of both microwave fields at their respective detunings. In the absence of trapping light and with only the $\sigma$ field present, the resonance frequency for the transitions between different dressed states $\ket{+}$ and $\ket{-}$ is $\omega_0 + \Delta_\sigma \pm \sqrt{\Omega_\sigma^2+\Delta_\sigma^2}$, where $\omega_0$ is the bare resonance frequency without dressing fields, and $\Omega_\sigma$ and $\Delta_\sigma = 2\pi \times 8.0$~MHz are the Rabi frequency and detuning for the $\sigma$ field. Measuring the resonant frequency allows us to accurately determine $\Omega_\sigma = 2\pi \times 8.0(1)$~MHz. Then, the $\pi$ dressing field with detuning $\Delta_\pi = 2\pi \times 10.75$~MHz mixes in additional states and causes energy shifts onto the aforementioned dressed states. From the resulting energy splittings, we extract the Rabi frequency of the $\pi$ microwave field of $\Omega_\pi = 2\pi \times 7.1(1)$~MHz. See Ref.~\cite{yuan2025upcoming} for more information.

To quantitatively characterize the ellipticity of the $\sigma$ microwave, we use an extended version of dressed-state spectroscopy. While dressed-state spectroscopy as described above is sensitive to microwave power, it does not provide information about polarization. To overcome this, we make additional use of the anisotropic polarizability of diatomic molecules~\cite{neyenhuis2012anisotropic, zhang2024dressed}. The idea is as follows: the $\sigma$ microwave field mixes the rotational state $\ket{0,0}$ with $\ket{1,-1}$, $\ket{1,0}$ and $\ket{1,1}$, each of which has a different optical polarizability at 1064 nm depending on the polarization angle of the optical field~\cite{zhang2024dressed}. By measuring the light shifts of the microwave-dressed molecules in an optical field with a variable linear polarization angle, we extract the state decomposition. As the state mixing depends on the Rabi frequency, the detuning, and the polarization (especially the ellipticity) of the microwave, this measurement allows us to reconstruct the polarization of the microwave field in the lab frame. The results are as follows: the microwave fields generated by the A-C and B-D antenna pairs are not perfectly linearly polarized, but instead have ellipticities of $30(3)\degree$ and $35(1)\degree$, respectively. The orientations of the polarization ellipses, $\hat{x'}$ and $\hat{y'}$, are rotated by $-14(3)\degree$ and $-11(4)\degree$ with respect to the lab frame coordinates $\hat{x}$ and $\hat{y}$ which are defined by the trap axes. The tilt of the microwave $\hat{x'} \hat{y'}$ plane is $\sim10(4)\degree$ from the trap $z$ axis for $\abs{\xi} < 10\degree$. With this method, we construct a model for the $\sigma$ field in the lab frame, which allows us to calculate the ellipticities in the microwave plane; the values are reported in the main text. The only remaining unknown parameter for the $\sigma$ field is the azimuthal Euler angle of the microwave plane, which contributes to the quoted uncertainty of the ellipticity. The minimum ellipticity of the $\sigma$ field achieved in this setup is $0.5(2) \degree$.

For the $\pi$ microwave field, the polarization is optimized by improving the evaporation efficiency as well as minimizing its projection onto the $\sigma$ microwave plane via reducing photon changing losses~\cite{yuan2025upcoming}. After optimization, the $\pi$ microwave contains $< \mr{sin}(1 \degree) \Omega_{\pi}$ of $\sigma^-$ component and $< \mr{sin}(4 \degree) \Omega_{\pi}$ of $\sigma^+$ component. A detailed discussion of the microwave calibration will be provided in Refs.~\cite{zhang2025upcoming, Stevenson2025upcoming}. 

{\bf Calculation of the interaction potential and scattering length.}
We compute the scattering length, dipolar length, and bound state positions from coupled-channels scattering calculations as described in Ref.~\cite{karman2025double}.

Dressing with two microwave fields prepares molecules in a field-dressed state $[u,v_\sigma,v_\pi]^\mathrm{T}$
 in the basis $\{|j,m,N_\sigma,N_\pi\rangle= |0,0,0,0\rangle, \cos\xi|1,1,-1,0\rangle+\sin\xi|1,-1,-1,0\rangle, |1,0,0,-1\rangle\}$,
 where $N_\sigma$ and $N_\pi$ indicate the number of photons in the two fields relative to a large reference number of photons, and $\xi$ characterizes the ellipticity of the $\sigma$ microwave field.

The long-range dipolar interaction between two molecules in the upper field-dressed state can be written as 
\begin{align*}
    U_{\mr{dd}}(\vec{r}) & = \frac{\hbar^2}{m} \Big[ a_{{\mathrm{d}0}} \left({{1 - 3 \cos ^2 \theta}} \right) + a_{{\mathrm{d}1}} {{\sqrt{3} \cos \varphi \sin 2 \theta}} \nonumber \\
    +& a_{{\mathrm{d}2}} {{\sqrt{3} \cos 2 \varphi \sin ^2 \theta}} \Big] r^{-3},
\end{align*}
where $\theta,\varphi$ are the polar angles of the intermolecular axis $\vec{r}$~\cite{karman2025double}.
For the microwave polarization considered here, the dipolar lengths of the individual components are given by:
\begin{align*}
a_{\mathrm{d}0} &= \frac{u^2(2v_\pi^2-v_\sigma^2)}{3} \frac{m d^2}{4\pi\epsilon_0 \hbar^2} ,\nonumber \\
a_{\mathrm{d}1} &= 0,\nonumber \\
a_{\mathrm{d}2} &= \frac{u^2v_\sigma^2}{\sqrt{3}} \frac{m d^2}{4\pi\epsilon_0 \hbar^2} \sin2\xi.
\end{align*}
For our microwave parameters, $a_{\mathrm{d}0} = 0$ and $a_{\mathrm{d}2}/\sin2\xi = 53,100~a_0$. The imperfection in the orthogonality between the two microwave fields leads to additional terms in the intermolecular interactions.
At compensation we estimate the residual dipolar interaction is given by $\abs{a_{\mr{d0}}} < 500~a_0$, $\abs{a_{\mr{d1}}} < 1,200~a_0$, and $\abs{a_{\mr{d2}}} < 600~a_0$.

The short-range repulsive interaction can be reasonably approximated by the dipolar interaction in second order \cite{deng2023effective}, which gives \cite{karman2025double}

\begin{widetext}
\begin{align*}
U_{\mr{vdW}}(\vec{r}) =& \frac{\hbar^2}{m} \Bigg[ {a_{6,0,0}^4} - {a_{6,2,0}^4} \left(3\cos^2\theta-1\right) - {a_{6,4,0}^4} \left(35\cos^4\theta-30\cos^2\theta+3\right) \nonumber \\
&- {a_{6,2,2}^4} \sin^2\theta \cos2\phi \sin2\xi - {a_{6,4,2}^4} \sin^2\theta \left(7\cos^2\theta-1\right)\cos2\phi \sin2\xi - {a_{6,4,4}^4} \sin^4\theta \cos4\phi \sin^22\xi \Bigg] r^{-6},
\label{eq:effpot2_new}
\end{align*}
\end{widetext}
where the van der Waals lengths are $a_{6,0,0}=3,200~a_0$, $a_{6,2,0}=1,400~a_0$,   $a_{6,4,0}=750~a_0$, $a_{6,2,2}=2,000~a_0$,   $a_{6,4,2}=1,300~a_0$, and $a_{6,4,4} = 1,900~a_0$. The resulting van der Waals interaction is almost isotropic as the $a_{6,0,0}^4$ term dominates. The imperfection in the orthogonality between the two microwave fields does not affect the short-range repulsive part significantly. The modifications to the van der Waals coefficients are on the order of $2\%$.

{\bf Roton instability for a uniform system under the ellipticity-tuned dipolar interactions.}
The dipolar interaction in this work has the form $U_{\mr{dd}}(\vec{r}) = B(\theta, \varphi)/r^\mr{3}$ with the prefactor given by $B(\theta, \varphi) = a_{\mr{d2}}(\xi) \hbar^2 {\mr{\sqrt{3} \cos 2 \varphi \sin ^2 \theta}}/m$. Using this, we obtain the Bogoliubov spectrum:
\begin{equation}
\nonumber
    \varepsilon_{\vec{k}} = \sqrt{\frac{\hbar^2 \vec{k}^2}{2m} \biggl\{ \frac{\hbar^2\vec{k}^2}{2m} + 2n_0\tilde{U}_\text{int}(\vec{k}) \biggr\} }\label{E_q}
\end{equation}
with 
\begin{align*}
    \tilde{U}_\text{int}(\vec{k}) = g + \tilde{U}_\mathrm{dd}(\vec{k})
    = g - \frac{4\pi}{3} B(\theta_k, \varphi_k) \nonumber \\
    = \frac{4\pi\hbar^2}{m} \left(a_{\mr{s}} - \frac{a_{\mr{d2}}(\xi)}{\mr{\sqrt{3}}} \cos 2 \varphi_k \sin ^2 \theta_k\right),
\end{align*}
 the Fourier transform of the total interaction potential. The roton instability for a uniform system happens when the Bogoliubov excitation energy reaches zero for a finite momentum, or equivalently, when $\tilde{U}_\text{int}(\vec{k}) < 0$. Therefore, for our system, it is expected to start at $\epsilon_{d2} = \left| a_{\mr{d2}}(\xi) \right| / a_{\mr{s}} = \sqrt{3}$, $\theta_k = 90\degree$, and $\varphi_k = 0\degree$ for positive ellipticity or $90\degree$ for negative ellipticity.

{\bf Model for \textit{in trap} density measurement via near-universal loss.}
We measure the \textit{in trap} molecular densities of the observed many-body states using decay induced by inelastic two-body loss. A similar method has been applied in magnetic atoms where three-body loss was used to extract densities~\cite{chomaz2016quantum, tanzi2019observation}. In order to work with a well-defined loss rate coefficient in our system, we utilize the near-universal two-body decay of unshielded molecules~\cite{julienne2011universal}.

After preparing the molecules into the desired many-body state, all microwave dressing fields are ramped down within 10 \textmu s, suddenly initiating inelastic collisions. This ramp time is non-adiabatic with respect to the molecular motion, but adiabatic with respect to the dressed state evolution, ensuring the resulting unshielded molecules are in the rotational ground state. The decay dynamics is observed for up to 3 ms. The molecule number is recorded using \textit{in situ} imaging to avoid number loss during time of flight. We then fit the number evolution with a loss model that is based on a local density approximation (LDA), assuming that molecules only collide with nearby molecules and do not rethermalize. This assumption is justified as the loss dynamics is faster than the thermalization time of around $(2 \pi) / \omega_z = 17$ ms. The number evolution can be numerically calculated as
\begin{equation}
\nonumber
\left\{ \begin{aligned}
    &N(t) = N_{\mr{deg}}(t) + N_{\mr{th}} = \int n_{\mr{deg}}(\vec{r}, t)d\vec{r} + N_{\mr{th}}, \\
    &\dot{n}_{\mr{deg}}(\vec{r}, t) = - \frac{\beta}{2} n_{\mr{deg}}(\vec{r}, t)^2,
\end{aligned} \right.
\end{equation}
where $N_{\mr{deg}}(t)$ is the total number of molecules in the quantum degenerate part of the gas; $N_{\mr{th}}$ is the molecule number in the thermal part and assumed to be constant (loss occurs dominantly in the degenerate part due to its higher density); $n_{\mr{deg}}(\vec{r}, t)$ is the density distribution of the degenerate part at spatial coordinate $\vec{r}$ and time $t$; $\beta$ is the near-universal two-body loss rate coefficient, which is determined experimentally (see below). The prefactor 1/2 accounts for correlations in the degenerate Bose gas~\cite{kagan1985effect}.

We fit the initial mean density $\left<n_0\right>$ and the thermal number $N_{\mr{th}}$ by assuming an isotropic Thomas-Fermi distribution for the initial density distribution of the degenerate part:
\begin{equation}
\nonumber
n_{\mr{deg}}(\vec{r}, t=0) =  \mr{max}\Bigg\{ \frac{7}{4}\left<n_0\right> (1-\abs{\vec{r}}^2/\sigma^2), ~0\Bigg\},
\end{equation}
where $\sigma = \left(\frac{15 N(t=0)}{14 \pi \left<n_0\right>}\right)^{1/3}$. 
Other fitting models, such as a single Gaussian or multiple Gaussian functions, give similar results within the overall uncertainty of the method. This robustness can be understood as the initial loss rate does not depend on the exact density profile, but only on the mean density $\left<n_0\right>$,  $\dot{N}_{\mr{deg}} / N_{\mr{deg}} = - \beta \left<n_0\right> /2$, at $t =0$.

\begin{figure}[]
  \centering
  \includegraphics[width = 8.6 cm]{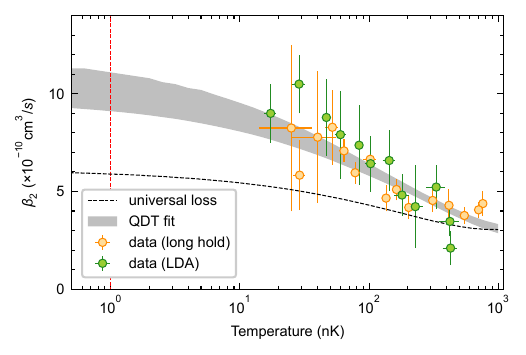}\\
  \caption{\textbf{Temperature dependence of the two-body loss rate coefficient for unshielded thermal NaCs ground-state molecules.} Green data points are obtained by fitting the decay for short hold times (up to 6 ms) using the LDA method. Orange data points are obtained by fitting the full temperature and number evolution over longer hold times (up to 100 ms). Horizontal error bars show the uncertainty in the temperature of the thermal molecular cloud; vertical error bars represent the uncertainty from fitting the cloud evolution. The gray shaded area shows the fitted theoretical result obtained from Quantum Defect Theory (QDT), representing the one sigma uncertainty from the fitted QDT loss parameters $\{y, \delta\}$. For comparison, the black dashed line shows the universal loss rate coefficient, which is incompatible with our measurements. The red vertical dotted line indicates the experimentally relevant temperature of $\sim 1$~nK. At this temperature, the extracted two-body loss rate coefficient is 1.0(1) $\times 10^{-9} \mr{cm}^{3}$/s.}
  \label{fig:SI4}
\end{figure}

{\bf Two-body loss rate coefficient for unshielded NaCs molecules.}
The two-body loss rate coefficient $\beta$ for bosonic molecules is known to have a weak temperature dependence~\cite{julienne2011universal}. The relevant temperature range of our degenerate samples is about 1 to 10~nK. To determine $\beta$ in this regime, we experimentally measure the loss rate coefficient for thermal clouds between 20 and 800 nK and then extrapolate to 1 nK (see Fig.~\ref{fig:SI4}). The experimental data is obtained via two independent measurement methods: (1) Using the LDA method as described above for a short hold times up to 6 ms, and (2) by holding the thermal cloud for up to $\sim$ 100 ms and fitting the complete temperature and number evolutions~\cite{yuan2025upcoming}. 

For the extrapolation, we follow the approach in Ref.~\cite{gregory2019sticky} and fit the experimental data to Quantum Defect Theory (QDT) calculations~\cite{idziaszek2010universal}, by numerically performing quantum mechanical scattering calculations for the rotational van der Waals interaction potential of unshielded molecules.
These calculations yield two linearly independent solutions which are used to impose arbitrary short-range boundary conditions parameterized by the QDT loss parameter $y$ and short-range phase shift $\delta$ using the method described in Ref.~\cite{karman2023resonances}.
We numerically propagate solutions using the Numerov algorithm between $r/r_6 = 0.3$ and 10, where $r_6$ is the van der Waals length, and for even partial waves up to $L=8$.
These calculations are repeated on a grid of collision energies,
and the resulting inelastic cross sections are thermally averaged over the Maxwell-Boltzmann distribution to obtain loss rate coefficients as a function of temperature for each pair of $\{y,~\delta \}$.
The parameter scans are done for $y$ between 0.01 and 1 in 21 steps, and the short-range phase shift between 0 and $\pi$ in 100 steps.
The resulting curves of loss rate coefficient versus temperature are then fit to the experimental data by minimizing chi-squared. Based on this, the extracted loss rate coefficient $\beta$ at 1 nK is 1.0(1)$\times 10^{-9} \mr{cm^3/s}$, for all possible pairs of $\{y,~\delta \}$ that are within one standard deviation from the experimental data. The value for $y$ is 0.46(19). Coefficients $\beta$ for those pairs are plotted in Fig.~\ref{fig:SI4}. We note that the unshielded loss is sufficiently rapid to neglect higher-order processes such as three-body recombination.

\clearpage

\begin{figure*}[]
  \centering
  \includegraphics[width = 18.0 cm]{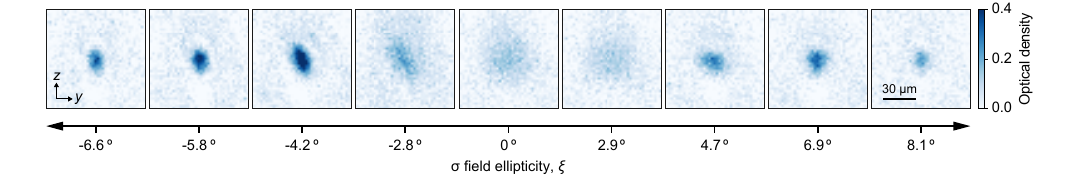}\\
  \caption{\textbf{Images of the molecular cloud taken from the $x$ direction.} Absorption images after 20 ms time-of-flight expansion using the same sequence as in Fig.~\ref{fig:2}. Images are taken with a different camera along the \textit{x} direction, the image plane corresponds to the $yz$ plane. Each image is an average of three individual shots. Scale bar, 30 \textmu m.}
  \label{fig:SI6}
\end{figure*}

\begin{figure*}[]
  \centering
  \includegraphics[width = 18.0 cm]{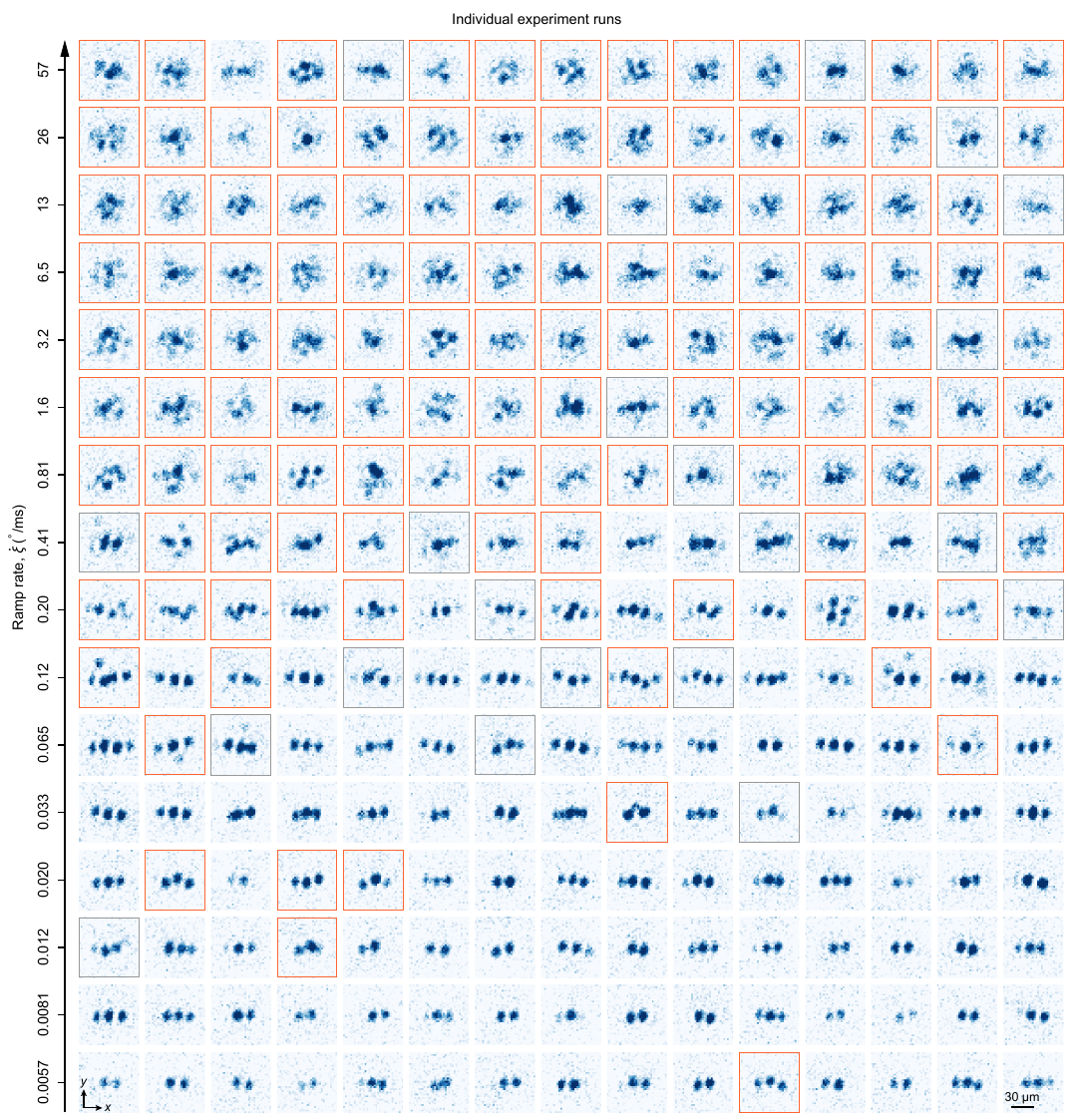}\\
  \caption{\textbf{Single-shot images of the molecular cloud after different interaction ramp rates.} Each row shows 15 shots for each ramp rate $\dot{\xi}$. The time of flight is 25 ms. Orange boxes mark images in which the formation of a 2D droplet array is observed; gray boxes mark images where the dimensionality cannot be clearly assigned and count towards the error bars in Fig.~\ref{fig:3}\textbf{c}; the remaining images show a 1D droplet array. Scale bar, 30 \textmu m.}
  \label{fig:SI2}
\end{figure*}

\begin{figure}[]
  \centering
  \includegraphics[width = 8.6 cm]{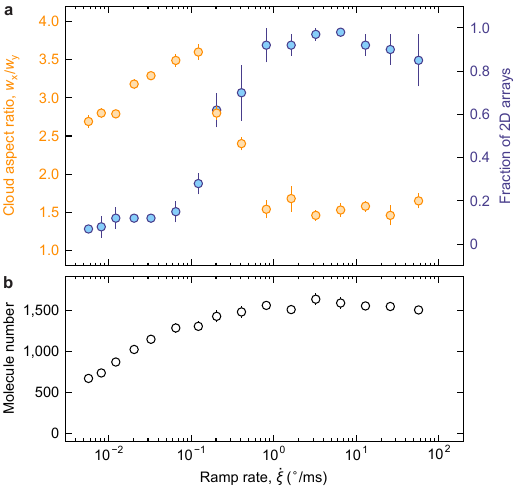}\\
  \caption{\textbf{Quantitative analysis of the adiabaticity of the interaction ramp.} \textbf{a}, Determining the crossover from 1D to 2D arrays. Orange data shows the aspect ratio $w_{\mr{x}} / w_{\mr{y}}$ of the averaged molecular cloud in the $xy$ plane, defined by the ratio of the full width half maximum along the $x$ and $y$ directions, obtained from a two-dimensional fit. Error bars show the uncertainty from fitting the width. For comparison, the blue data (identical to Fig.~\ref{fig:3}\textbf{c}) shows the fraction of images exhibiting 2D arrays, as illustrated in Fig.~\ref{fig:SI2}. Error bars represent the uncertainty in determining the dimensionality of the array. Both analysis methods agree with each other, indicating a crossover at around $\dot{\xi} = 0.3 \degree$/ms. \textbf{b}, Average number of molecules after the interaction ramp with rate $\dot{\xi}$. The number remains almost constant across the crossover region, excluding the possibility that the transition from 1D to 2D structures is caused by a changing molecule number. Each data point is obtained by fitting the averaged image of 30 shots. Error bars denote the uncertainty in the fitted molecule number.}
  \label{fig:SI7}
\end{figure}

\begin{figure}[]
    \centering
    \includegraphics[width = 8.6 cm]{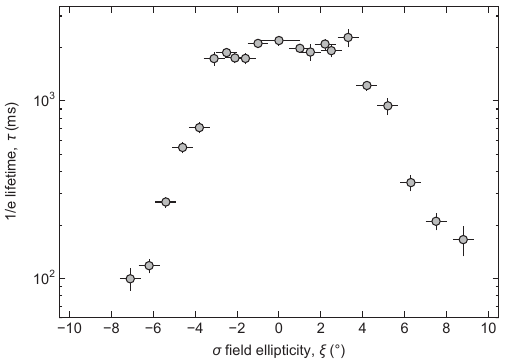}\\
    \caption{\textbf{Lifetime in the BEC and droplet phases.} Measured 1/e lifetime of the molecules gas in the trap after ramping the ellipticity to $\xi$ within 200 ms. Each data point is obtained by fitting an exponential decay curve to the evolution of total molecule number as a function of the hold time. Horizontal error bars show the uncertainty in the ellipticity calibration; vertical error bars show the uncertainty in the fitted lifetime.
    } 
    \label{fig:SI8}
\end{figure}

\end{document}